# Scale-free wave localization at topological multicriticality

Zhijie Xue[1,7], Yuan Tian[1,7], Nanyu Chen[1,7], Xiujuan Zhang[1,*], Tao Li[1,*], Shuang Zhang[2,3,4,5,6,*], Fuxin Guan[1,*]

[1]National laboratory of Solid State Microstructures, Jiangsu Key laboratory of Artificial Functional Materials, Collaborative innovation center of advanced micro-structures, College of engineering and Applied Sciences, Nanjing University, Nanjing 210093, China

[2]New Cornerstone Science Laboratory, Department of Physics, The University of Hong Kong, Hong Kong 999077, China

[3]State Key Laboratory of Optical Quantum Materials, The University of Hong Kong, Hong Kong 999077, China

[4]Department of Electrical and Electronic Engineering, The University of Hong Kong, Hong Kong 999077, China

[5]Quantum Science Center of Guangdong-Hong Kong-Macao Great Bay Area, 3 Binlang Road, Shenzhen, China

[6]Materials Innovation Institute for Life Sciences and Energy (MILES), HKU-SIRI, Shenzhen, People's Republic of China

[7]These authors contributed equally

[*]Email: xiujuanzhang@nju.edu.cn; taoli@nju.edu.cn; shuzhang@hku.hk; guanfuxin@nju.edu.cn

**Abstract:**

Spatial wave localization at boundaries is conventionally restricted to exponential decay. Here, we observe a scale-free boundary mode exhibiting anomalous polynomial decay, highlighted by strictly linear spatial envelopes, driven by topological multicriticality in acoustic metamaterials. We establish a universal geometric bulk-boundary correspondence rooted in singularity theory by mapping $n$-th order topological multicriticality onto Arnold's $A_n$ catastrophe. We prove that defectiveness in the spatial transfer matrix enforces a $(n-1)$-th power spatial decay of boundary modes. At a topological tri-critical point, the singularity of winding function manifests as an $A_2$ fold catastrophe. This dictates a strictly linear spatial decay, marking a sharp departure from the conventional exponential scaling paradigm. Using acoustic metamaterial platforms tuned to a tricritical point, we experimentally demonstrate this linear decay envelope in 1D, and further generalize it to two-dimensional systems to realize scale-free corner modes with simultaneous linear decay along orthogonal directions. While existing critical topology frameworks rely on entanglement spectra to diagnose mode existence in gapless regimes, our work bridges catastrophe theory with topological matter at multicriticalities, advancing the field from abstract mode identification to deterministic real-space wavefunction engineering.

## Introduction

Spatial wave localization at interfaces play a crucial role in wave physics, governing the confinement, transport, and manipulation of energy across photonic[1–6], acoustic[7–11], and quantum platforms[12,13]. The envelope of an interfacial mode is dictated by the dispersion relation of the host medium[14]. Within allowed passbands, wavefunctions extend as Bloch states with delocalized envelopes; within spectral bandgaps, boundary modes decay exponentially into bulk. While conventional wave confinement properties are widely exploited for waveguiding, tunneling and energy trapping, a fundamental open question in wave physics is whether spatial localizations can transcend conventional exponential decay paradigm[15–18].

As a cornerstone of topological physics, the bulk-boundary correspondence (BBC) counts localized boundary modes by mapping them to global topological invariants[19–27], such as integer Chern numbers[20,21], defined over closed Brillouin zones. Owing to the topological robustness of these integer invariants under smooth perturbations, physical interfaces separating distinct topological domains are guaranteed to harbor topologically protected boundary modes[28–36]. But this conventional paradigm encounters a fundamental breakdown at topological phase transitions and quantum critical points, where the closing of the bulk gap renders conventional invariants ill-defined[37,38]. Recent breakthroughs resolve this gap-dependent paradigm by demonstrating the existence of critical topology and using the entanglement spectrum to successfully diagnose topological boundary modes within gapless bulk spectra[39–45]. Nevertheless, both approaches remain confined to a mere mode counting tool that tacitly assumes conventional exponential wave decay[46–48] , thereby failing to capture their exact spatial envelopes.

Here, we discover an anomalous boundary mode with linear spatial decay at topological multicriticalities and establish a geometric BBC to predict this behavior. By reformulating BBC through the lens of catastrophe theory[49–52], we classify root coalescence of winding function via Arnold's ADE classification[53] to dictate exact wave profiles. Extending SSH lattice by incorporating long-range intercell hoppings, the Hamiltonian acquires higher-order Fourier harmonics, driving $n$-fold root coalescence of the winding function at multicritical points[54–58]. We reveal that an $n$-th order Lifshitz degeneracy in the topological phase diagram maps precisely onto an $A_n$ catastrophe. This geometric singularity collapses the transfer matrix of wavefunction into a defective Jordan canonical form, imprinting an exact $(n-1)$-th power algebraic spatial scaling law directly onto the boundary mode.

Specifically, the $A_2$ fold catastrophe forces a strictly linear spatial profile, marking a departure from conventional exponential localization. Using a 1D acoustic metamaterial platform tuned to the intersection of three distinct topological winding phases, we experimentally observe this anomalous linear scaling law. Furthermore, we generalize this framework to 2D higher-order topological acoustic metamaterials tuned to the tricritical convergence of three phases characterized by multipole chiral numbers. We experimentally observe scale-free corner modes displaying simultaneous linear decay along both orthogonal directions. Our work establishes a unified geometric bulk boundary correspondence between catastrophe manifolds and real space wavefunction engineering, providing a universal blueprint to tailor anomalous wavefunctions across classical and quantum systems.

**Result:**

We start with an investigation of zero mode in SSH model system with long-range intercell hopping up to the $n$-th order, $H_{PBC}(k) = d_x(k)\sigma_x + d_y(k)\sigma_y$, where the auxiliary quantities $d_x(k)$ and $d_y(k)$ define the winding function $w(k) = d_x(k) - id_y(k) = \sum_{m=0}^{n} t_m e^{\mathrm{i}mk}$. Introducing the conformal mapping $z = e^{\mathrm{i}k}$ transforms the winding function into $w(z) = \sum_{m=0}^{n} t_m z^m$. The corresponding open Hamiltonian takes the form as $H_{OBC} = \sum_{m=0}^{n} \sum_{j=1}^{N_B} t_m \left( |B_j\rangle\langle A_{j+m}| + h.c. \right)$, where $m \le N_A - j$ and $N_A = N_B + 1$, as schematically depicted in Fig. 1(a). The general eigenstate is written as $|\psi\rangle = \sum_j^{N_A} a_j |A_j\rangle + \sum_j^{N_B} b_j |B_j\rangle$. Under chiral symmetry, the zero-energy wavefunctions on sublattice B vanish ($b_j = 0$). Consequently, the Schrodinger equation on sublattice A reduces to an $n$-th order linear homogeneous recursion relation $\sum_{m=0}^{n} t_m\, a_{j+m} = 0$. Defining the state vector $\mathbf{A}_j = (a_{j+n-1},\ a_{j+n-2},\ \cdots,\ a_j)^T$, the propagation between the $j$-th and ($j$+1)-th state vector is governed by the $n \times n$ transfer matrix $\mathbf{A}_{j+1} = \mathbf{T}\mathbf{A}_j$,

$$\mathbf{T} = \begin{pmatrix} -\frac{t_{n-1}}{t_n} & -\frac{t_{n-2}}{t_n} & \cdots & -\frac{t_1}{t_n} & -\frac{t_0}{t_n} \\ 1 & 0 & & 0 & 0 \\ 0 & 1 & & 0 & 0 \\ \vdots & \vdots & \ddots & \vdots & \vdots \\ 0 & 0 & \cdots & 1 & 0 \end{pmatrix} \tag{1}$$

The corresponding characteristic equation $P(\lambda) = \det(\mathbf{T} - \lambda\mathbf{I}) = 0$ is expressed as,

$$P(\lambda) = \sum_{m=0}^{n} t_m \lambda^m = 0 \tag{2}$$

At a $n$th order critical point, all $n$ eigenvalues collapse at either $\lambda = 1$ or $-1$. We here choose $\lambda = -1$ for investigation, and thus the characteristic polynomial can be reduced to $P(\lambda) = t_0(\lambda + 1)^n = 0$. Matching coefficients of Eq. (2) via binomial expansion determines the critical hopping parameters, derive

$$t_m = \binom{n}{m} t_0 \tag{3}$$

Crucially, the winding function $w(z) = 0$ at this critical condition is mathematically identical to the OBC characteristic equation $w(z) = \det(\mathbf{T} - z\mathbf{I}) = P(z) = 0$, bridging the bulk topological property to the OBC transfer matrix of zero mode.

Substituting Eq. (3) into $H_{PBC}$, we derive the band structures from $n$ = 2 to 5 in Fig. 1(b). The two bands degenerate at $K$ point, where the eigen energy holds properties of $|E| \propto |(k - \pi)|^n$. The corresponding zero eigenstates are depicted by the grey lines in Figs. 1(c)–1(f), respectively. Notably, the spatial profiles steepen rapidly with higher critical order $n$.

To understand this spatial scaling property, we analyze the defective structure of the transfer matrix Eq. (1) at the critical condition Eq. (3) directly in its Jordan normal form $\mathbf{T} = \mathbf{P}\mathbf{J}_n\mathbf{P}^{-1}$, where $\mathbf{J}_n = -\mathbf{I}_n + \mathbf{N}_n$ represents an $n \times n$ Jordan block with $\mathbf{N}_n$ being a nilpotent matrix satisfying $\mathbf{N}_n^n = 0$. The transformation matrix $\mathbf{P} = (\mathbf{v}_1, \mathbf{v}_2 \cdots \mathbf{v}_n)$ is constructed from the Jordan basis . The propagation of the *j*th state vector is evaluated using the high-power transfer matrix $\mathbf{A}_j = \mathbf{T}^{j-1}\mathbf{A}_1 = \mathbf{P}\mathbf{J}_n^{j-1}\mathbf{P}^{-1}\mathbf{A}_1$, yielding the exact spatial eigenstate distribution from the end of the lattice:

$$|a_j| = \binom{N_A + n - j - 1}{n - 1} |a_{N_A}|. \tag{4}$$

When $N_A$ is large enough, Eq. (4) can be simplified as $|a_j| \sim (N_A - j)^{n-1}$, with the detailed analytical derivation provided in Supplementary section I. Using Eq. (4) to evaluate the field distribution, we derive the results for *n*=2 to 5 as depicted by the red lines in Figs. 1(c)-1(f), which matches perfectly with the grey lines. Specifically, for *n*=2, the wavefunction exhibits a perfect linear scaling, fundamentally distinct from conventional exponential localization.

Next, we employ Arnold's catastrophe classification to reveal the universal geometric origin of this unconventional scaling law. We begin by recognizing that right at the unperturbed critical point, the system's equilibrium states correspond to the roots of a topological potential derivative: $\left.\frac{\partial V_p}{\partial \lambda}\right|_{unperturbed} = w(\lambda) = P(\lambda) = 0$. To track how these states evolve under small parameter deviations away from strict multicriticality, we introduce a set of perturbation control parameters $\mathbf{u} = (u_1, u_2, \cdots, u_{n-1})$ to parametrize the detuning away from the critical point. Centering our coordinate system around the n-fold degenerate root at $\lambda = -1$ via the shift $\lambda \to \lambda - 1$, the perturbed potential derivative is represented by superimposing the unperturbed baseline $P(\lambda - 1)$ with a multi-variable Taylor expansion in parameter space $\frac{\partial V_p}{\partial \lambda} = P(\lambda - 1) + \sum_{k=0}^{n-2} \frac{\partial^{(n-k)} P(\lambda-1)}{(n-k)!\partial \lambda^{(n-k)}} u_{k+1} = 0$. Right at the *n*-th order defective point where $P(\lambda - 1) = \lambda^n$, evaluating these higher-order derivative terms simplifies the Taylor series into a clean, canonical polynomial known as the universal unfolding potential function:

$$\frac{\partial V_p}{\partial \lambda} = \lambda^n + \sum_{k=0}^{n-2} u_{k+1}\lambda^k \tag{5}$$

It concludes from Eq. (5) that catastrophe theory rigorously establishes the mathematical equivalence between an $n$-fold characteristic root and an $A_n$ singularity. Utilizing Eq. (5), we plot the equilibrium manifolds for the Fold ($A_2$), Cusp ($A_3$), Swallowtail ($A_4$), and Butterfly ($A_5$) catastrophes in Figs. 1(g)–1(j), respectively. The red dots pinpoint the critical singularities at $\mathbf{u} = 0$, where corresponds to the condition of Eq. (3). Ultimately, this $A_n$ catastrophe encoded in the characteristic of the transfer matrix dictates the universal spatial scaling $|a_j| \propto (N_A - j)^{n-1}$.

The above results have clarified the correspondence from the winding function to spatial envelope via the Arnold's catastrophe. Next, we focus on the topological behavior and its spatial scaling of boundary mode classified by $A_2$ catastrophe. Evaluating the topological winding number, we construct the topological phase diagram as a function of $t_0$ and $t_2$ with fixed $t_1 = 1$, as illustrated in Fig. 2(a). The parameter space hosts three distinct topological phases characterized by winding number $w_N = 0,1,2$. These phases are separated by gapless phase boundaries that intersect at a tri-critical Lifshitz point [red star in Fig. 2(a)]. The corresponding energy and winding spectra are displayed in Fig. 2(b). Enforced by chiral symmetry, the non-zero eigenstates appear in pairs, and the odd lattice gives a zero mode. For comparison with the open boundary spectrum, the insets present the winding trajectory of the auxiliary field $(d_x, d_y)$. The auxiliary field traces a characteristic cardioid curve whose tip precisely touches the origin, where this tangency marks the bulk gap closing and manifests as a $A_2$ fold catastrophe.

To assess the topologically critical property near this Lifshitz point, we investigate the winding root dynamics though the trajectory evolution of the three phase transition boundaries, as shown in Fig. 2(c). The three critical boundaries exhibit fundamentally different critical topologies, which are related to the movement of the roots (blue dots) relative to the unit circle $|z| = 1$ (grey circles). Along the red boundary, one zero remains enclosed inside the unit circle while the second zero locates at $z=-1$ (bottom right), guaranteeing its non-trivial criticality. Conversely, along the blue line, a single zero approaches $z=-1$ from outside the unit circle with no root enclosed (bottom-left circle), marking a topologically trivial criticality. Along the green line, two complex conjugate zeros migrate directly along the unit circle toward $z=-1$ (top circle). Ultimately, all trajectory paths converge at the central tri-phase point (red star), where the two zeros coalesce strictly at $z=-1$ on the unit circle. All these analysis matches with Ref.[39,40] .

Here we test the robustness of this linear wave mode. By increasing the total lattice sites from 39 to 119, we plot wavefunctions in Fig. 2(d) and find that the linear decay is independent of lattice size. The wavefunction evolution as the hopping parameters sweeping from the red to blue lines [Fig. 2(a)] is depicted in Fig. 2(e). Notably, the spatial profile smoothly transforms from concave to convex, demonstrating that the linear scaling represents a robust, continuous variation rather than an abrupt, discontinuous feature.

We design acoustic metamaterials to realize this anomalous linear scaling, and begin by a 1D array of coupled acoustic cavities [Fig. 3(a)]. The sample consists of 19 coupled acoustic cavities, each cavity supports a first-order resonance at approximately 4.01 kHz, which serves as the effective onsite energy of the acoustic tight-binding model. The couplings between the cavities are implemented by connecting channels and are characterized by the effective hopping amplitudes $t_0$, $t_1$, and $t_2$. To experimentally examine the critical behaviors discussed above, we fabricate three representative samples corresponding to the three marked points in the phase diagram of Fig. 2(b). For the tricritical Lifshitz point, the effective couplings are chosen as $t_1=-100$ Hz, $t_0=-50$ Hz, and $t_2=-50$ Hz. For the critical position at the blue dot, we employ $t_1=-100$ Hz, $t_0=-70$ Hz, and $t_2=-30$ Hz, whereas for the red circle we set $t_1=-100$ Hz, $t_0=-30$ Hz, and $t_2=-70$ Hz. The three measured dispersions show good agreement with the theoretical calculations for all three samples [Figs. 3(b)]. In particular, the tricritical sample exhibits a quadratic band touching at the K point, whereas the other two cases display linear band crossings.

The corresponding eigenfrequency spectra are displayed in Fig. 3(c), where the tight-binding calculations agree well with the full-wave simulations for all three parameter sets. In each case, the mode highlighted by the red circle is selected for further analysis, and the corresponding pressure-field simulations are presented in Fig. 3(d), providing a direct visualization of the spatial characteristics of these critical modes. For the first and third cases, the waveforms exhibit bulging and concave profiles, respectively. In contrast, at the tricritical Lifshitz point, the mode extends throughout the finite chain and develops a linear spatial envelope. Each measured site-resolved pressure dots of these three cases in Fig. 3(e) shows good agreement with the corresponding theoretical profile, respectively.

To demonstrate the universality of this catastrophe-induced scaling, we extend our framework to 2D lattices [Fig. 4(a)]. The 2D Hamiltonian with chiral symmetry is expressed as $H_{2D}(\boldsymbol{k}) = \begin{pmatrix} 0 & h_{total}(\boldsymbol{k}) \\ h_{total}^{\dagger}(\boldsymbol{k}) & 0 \end{pmatrix}$, where $h_{total}(\boldsymbol{k}) = l(\boldsymbol{k}) + g(\boldsymbol{k})$ incorporates both nearest- and next-nearest-neighbor couplings, $l(\boldsymbol{k}) = \begin{pmatrix} t_0 + t_1 e^{-ik_x} & -t_0 - t_1 e^{-ik_y} \\ -t_0 - t_1 e^{ik_y} & -t_0 - t_1 e^{ik_x} \end{pmatrix}$ and $g(\boldsymbol{k}) = \begin{pmatrix} t_2 e^{-2ik_x} & -t_2 e^{-2ik_y} \\ -t_2 e^{2ik_y} & -t_2 e^{2ik_x} \end{pmatrix}$. Calculating the multipole chiral number (MCN) constructs a topological phase diagram harboring a tricritical junction, with the three MCN numbers are $(N_{xy} = 4, 1, and\ 0)$, respectively[59–62].

We analytically evaluate the zero-energy mode distribution at the tricritical condition $(2t_0 = t_1 = 2t_2)$ by solving $H_{OBC}\psi = \begin{pmatrix} 0 & h_{total} \\ h_{total}^{\dagger} & 0 \end{pmatrix}\psi = 0$, where $\psi = \begin{pmatrix} \psi_A \\ \psi_B \end{pmatrix} = (\psi_1, \psi_2, \psi_3, \psi_4)^T$ represents the four site amplitudes per unit cell [inset of Fig. 4a]. The chiral block structure enforces $h_{total}^{\dagger}\psi_A = 0$ and $h_{total}\psi_B = 0$. Explicit calculation yields $\dim\ker\left(h_{total}^{\dagger}\right) = 1$ and $\dim\ker(h_{total}) = 0$. Thus, $h_{total}^{\dagger}$ admits a unique zero mode, up to an overall normalization, satisfying $\psi_2 =$ $\psi_3 = \psi_4 = 0$, whereas $h_{total}$ has no zero energy contribution. Consequently, the 2D problem decouples into two orthogonal 1D differential conditions for the mode $\psi_1(x, y)$ : $-(t_0 + t_1 e^{ik_x} + t_2 e^{2ik_x})\psi_1(x, y) = 0$ and $(t_0 + t_1 e^{ik_y} + t_2 e^{2ik_y})\psi_1(x, y) = 0$. At the tricritical singularity, this operator separability allows each spatial dimension to independently recover the 1D $A_2$ catastrophe condition, yielding an exact 2D spatial envelope:

$$\psi_1(x, y) = (N_x + 1 - x)(N_y + 1 - y) \tag{6}$$

where $N_x$ and $N_y$ indicate the total lattice number in x and y directions. The detailed mathematical derivations are provided in Supplementary Material section II.

We design the acoustic metamaterial lattices with the schematic displayed in Fig. 4(b). By exciting the cavity at the left bottom using the synthesized complex frequency approach[63–65], we obtain the simulated and measured pressure distributions in Fig. 4(c), with the measured results match perfectly with the simulations. To illustrate the linear decay, pressure profiles along two representative trajectories [pink and yellow lines in Fig. 4(b)] are plotted in Fig. 4(d), using color coding consistent with the trajectories. The measured results agree well with the theoretical linear behavior. These results confirm the bilinear localization predicted by Eq. (7) and demonstrate the extension of catastrophe-governed linear spatial decay from 1D to 2D.

**Summary**

We report here, to our knowledge, the first experimental observation of scale-free boundary modes exhibiting linear decay profiles, together with an extended bulk–boundary correspondence that quantitatively predicts their envelopes. By mapping multicritical gap-closing singularities onto Arnold's catastrophe manifolds in acoustic metamaterials, we reveal that catastrophe of topological criticalities quantitatively dictates boundary wave envelopes, advancing topological physics from discrete mode counting to deterministic envelope engineering. Furthermore, we generalize this geometric scaling to 2D lattices and experimentally realize a higher-order topological acoustic metamaterial tuned to the tricritical convergence of three distinct multipole chiral phases. We observe scale-free corner modes that exhibit simultaneous linear spatial decay along both orthogonal axes, demonstrating that the $A_2$ catastrophe governs higher-order corner states independently along both directions. Our work bridges singularity theory with topological physics, providing a universal framework to engineer custom wavefunctions with broad implications.

**Acknowledgement**

We sincerely thank the fruitful discussion with Prof. Guancong Ma and Dr. Xulong Wang from Hong Kong Baptist University.

## Figures

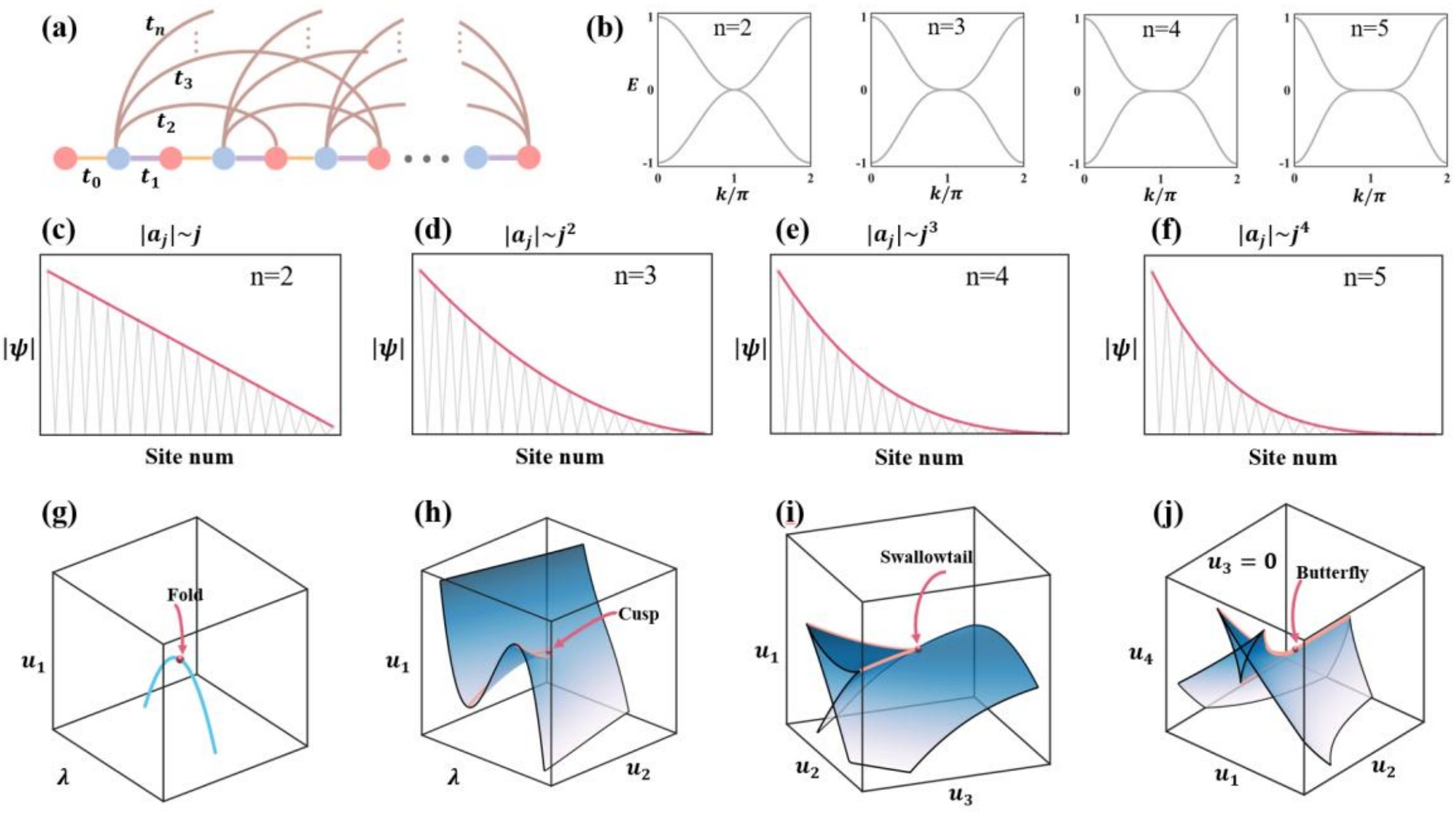

**Figure 1.** Anomalous wave localization and their correspondence with Arnold's catastrophe representations. (a) Schematic of the extended one-dimensional SSH chain with hopping parameters are displayed inset. (b) Band structures at *n*th order multicriticality. (c–f) Zero-energy wavefunctions at multicritical singularities with orders from 2 to 5. The site-resolved amplitudes are shown in gray and their envelopes in red. (g–j) Canonical unfolding geometries associated with these multicriticalities: fold, cusp, swallowtail, and butterfly catastrophes. The variable $\lambda$ represents the eigenvalue and $u_i$ are the unfolding parameters. Red points mark the singular loci, while the blue manifolds show the corresponding equilibrium surfaces.

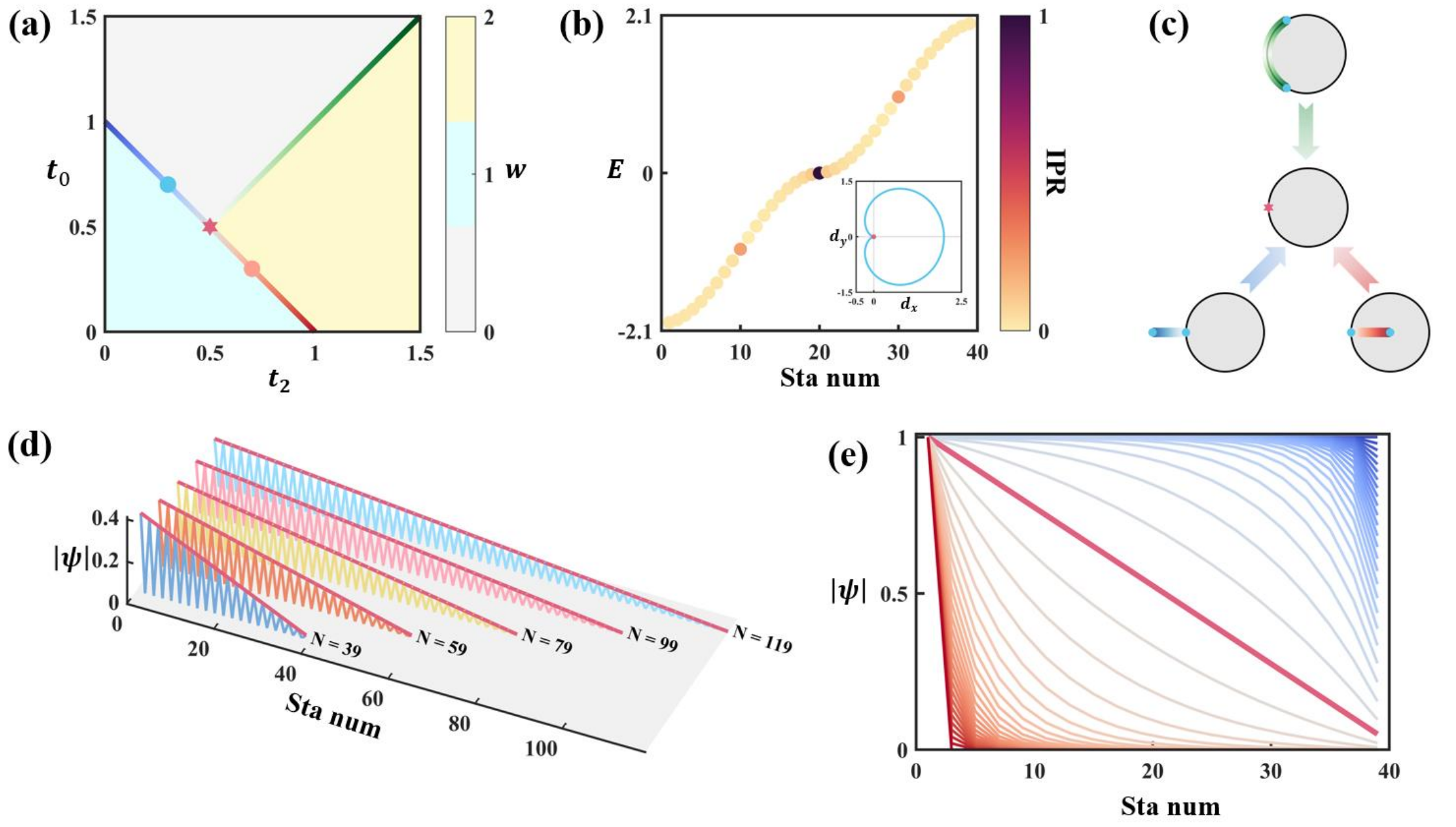

**Figure 2.** Topological behaviors and linear spatial scaling classified by $A_2$ catastrophe. (a) Topological phase diagram in the $(t_2,\ t_0)$ plane with $t_1 = 1$. Three topological phase boundaries, highlighted in blue, green and red, converge at the tricritical Lifshitz point marked by the red hexagon. (b) Eigenenergy spectrum at the red hexagon position. The color scale denotes the inverse participation ratio (IPR). Insets show the corresponding band structure and the winding trajectory of the $d(k)$ vector. (c) Schematic of the winding root positions of Eq. (6) and the unit circle. The colors correspond to that of the phase boundaries in (a). (d) Spatial profiles of the zero-energy state for finite chains of different lengths from 39 to 119. (e) Evolution of the zero-energy wavefunction envelopes upon approaching the topological tricritical point from blue and red critical lines in (a).

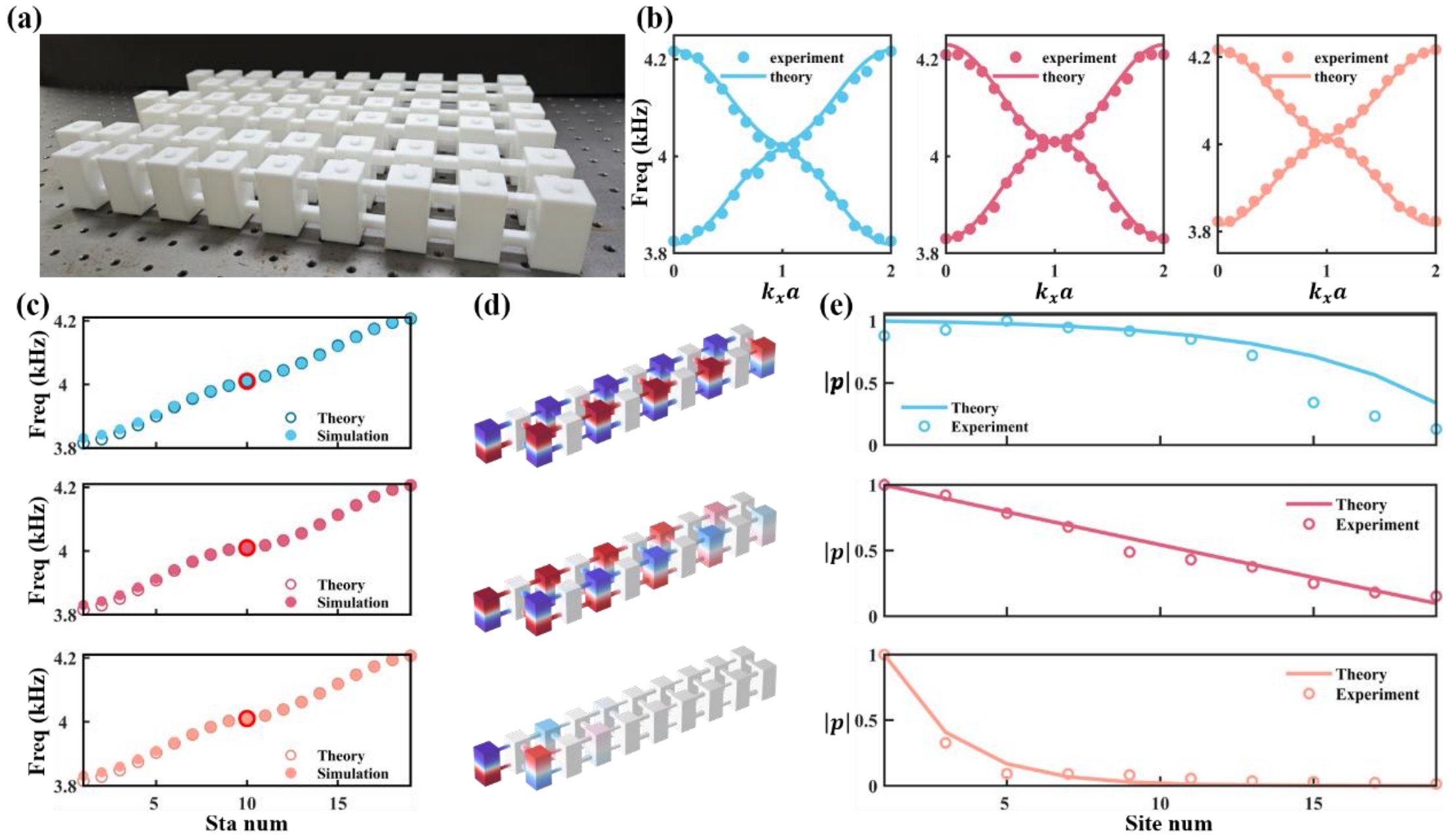

**Figure 3.** Experimental observation of the linear spatial scaling at the topological tricritical point. (a) Photograph of the three distinct 19-site acoustic chains, whose hopping parameters correspond to the three colored points marked in Fig. 2(b). (b) Measured band structures for the three cases compared with theoretical results. The tricritical point exhibits a quadratic band touching, while the other two cases show linear crossings. (c) Finite-chain eigenfrequency spectra from theory and full-wave simulations. (d) Simulated pressure-field distributions of the zero-modes. (e) Measured site-resolved amplitudes of these three cases compared with theoretical results.

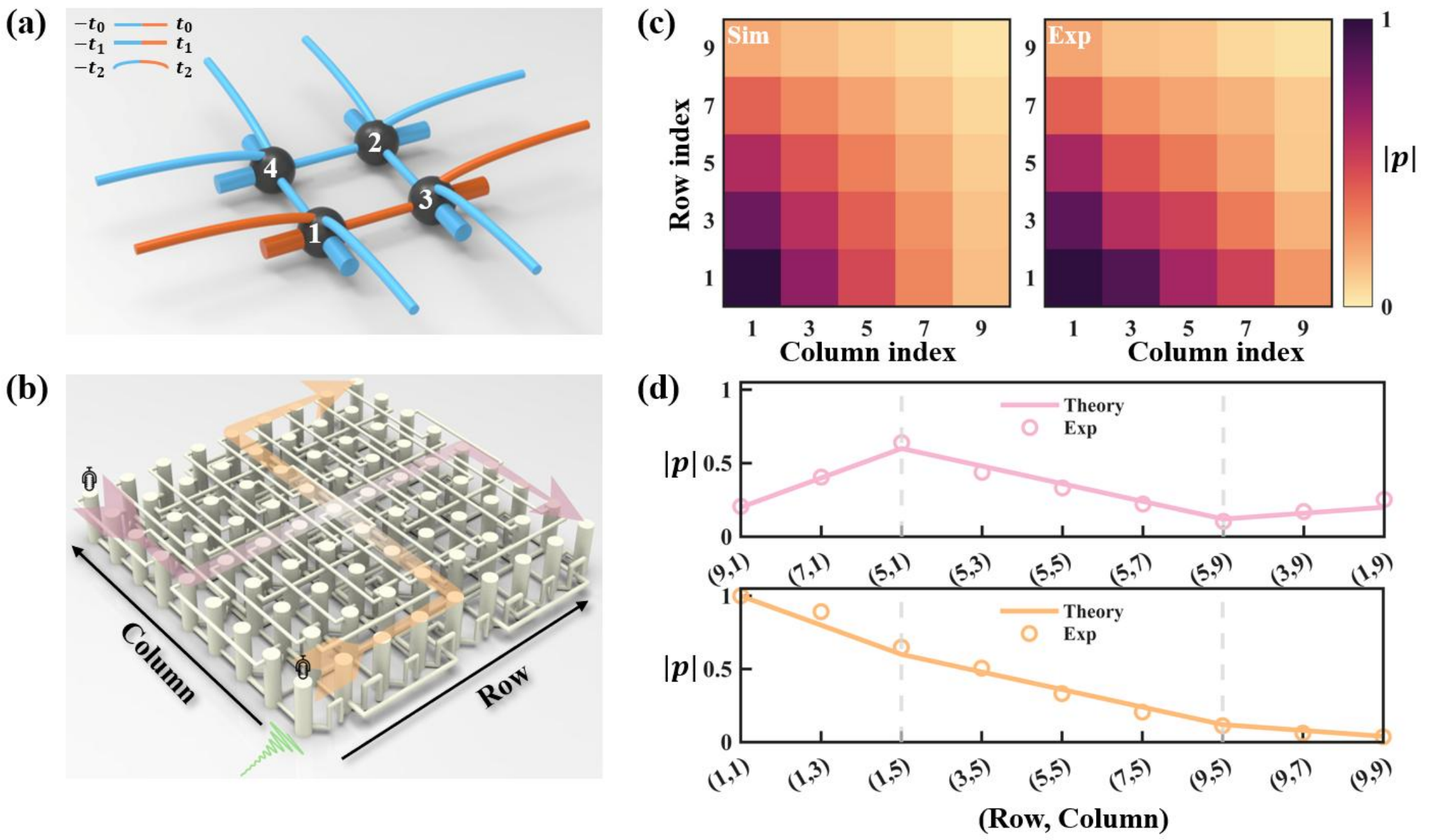

**Figure 4.** Two-dimensional scale-free corner modes at tricriticality, exhibiting simultaneous linear decay along both dimensions. (a) Schematic of the 2D unit cell with hopping parameters and labels are depicted in the inset. (b) Schematic of the 2D acoustic metamaterial lattices, comprising 9 lattice sites along each direction. (c) Measured and simulated pressure distributions of the corner state on the odd sublattice sites. (d) Measured pressure profile extracted along the two cut lines in (b), the vertical grey dashed lines separate different parts.